\documentclass[showpacs,aps,prb,twocolumn,groupedaddress]{revtex4}
\usepackage{graphicx}
\usepackage[dvipdfm]{hyperref}
\hypersetup{colorlinks=true,linkcolor=blue,
  implicit=true,breaklinks=true,pagebackref=true,backref=true,
bookmarks=true,bookmarksnumbered=true,hyperfootnotes=true,debug=true,
  naturalnames=false,citecolor=blue,pdfview=FitH,pdfstartview=FitH,hyperindex=true}
\usepackage{amsmath}
\usepackage{amssymb}
\begin{document}

\title{From local moment to mixed-valence regime in Ce$_{1-x}$Yb$_x$CoIn$_5$ alloys} 

\author{Y. P. Singh$^\star$, D. J. Haney$^\star$, X. Y. Huang$^\star$, I. K. Lum$^\ddagger$, B. D. White$^\ddagger$, M. Dzero$^\star$, M. B. Maple$^\ddagger$ and C. C. Almasan$^\star$}
\affiliation{$^\star$Department of Physics, Kent State University, Kent, Ohio, 44242, USA, \\
$^\ddagger$ Department of Physics, University of California at San Diego, La Jolla, CA 92903, USA}

\date{\today}

\begin{abstract}
We investigated the onset of the many-body coherence in the $f$-orbital single crystalline alloys Ce$_{1-x}$Yb$_x$CoIn$_5$ ($0 \leq x \leq 0.775$) through thermodynamic and magneto-transport measurements.
Our study shows the evolution of the many-body electronic state as the Kondo lattice of Ce moments is transformed into an array of Ce impurities. Specifically, we observe a crossover from the predominantly localized Ce moment regime to the predominantly itinerant Yb $f$-electronic states regime for about 60$\%$ of Yb doping. Our analysis of the residual resistivity data unveils the presence of correlations between Yb ions, while from our analysis of specific heat data we conclude that for $0.65 \leq x \leq 0.775$, Yb $f$-electrons strongly interact with the conduction electrons while the Ce moments remain completely decoupled.  
The sub-linear temperature dependence of resistivity across the whole range of Yb concentrations suggest the presence of a unconventional scattering mechanism for the conduction electrons. 
\end{abstract}

\maketitle

\section{Introduction}
In recent years, there has been a remarkable upturn in experimental work reporting the formation of highly unusual correlated many-body states in materials with partially occupied $f$-orbitals. Examples include observation of the topologically protected metallic surface states in canonical Kondo insulator SmB$_6$,\cite{exp1smb6,ARPES_Zahid,SurfaceDoping} violation of the Wiedmann-Franz law in YbRh$_2$Si$_2$,\cite{WiedemannFranzViolate} intrinsic non-Fermi liquid behavior and unconventional superconductivity in $\beta$-YbAlB$_4$,\cite{betaYball} rotational symmetry breaking inside the hidden-order phase in URu$_2$Si$_2$,\cite{MydoshReview,Okazaki} and cooperative states in superconducting alloys Ce$_{1-x}$Yb$_x$CoIn$_5$.\cite{LiShu} Despite many years of experimental and theoretical research, however, many of these phenomena---with the possible exception of metallic states with Dirac spectrum in SmB$_6$---are still poorly understood and are often a subject of controversy. 

What is common to all these phenomena is that they are driven not only by the strong local correlations between the $f$-electrons, but also by the non-local interactions between the conduction and $f$-electrons.\cite{PiersReview}
In particular, large enough repulsion between the $f$-electrons may lead to a significant energy separation between the two ionic configurations $f^n$ and $f^{n+1}$, thus leading to the formation of the local magnetic moments.\cite{Stewart} At the same time, in other cases, even in the presence of strong Coulomb interactions, the energy of the two ionic configurations becomes comparable, so that the local moments do not form and the $f$-ions are in an intermediate valence state.\cite{Brian1, Brian2, Chandra} How unconventional superconductivity emerges in the $f$-orbital materials that appear to be either in the local moment or mixed valence regime still remains one of the most fundamental problems.

One way to tune the relevant energy scales in order to study the physics of the crossover from Kondo-lattice to Kondo-impurity behavior is to introduce substitutional disorder on the $f$-electron sites. In the case of the unconventional superconductor CeCoIn$_5$---a member of the family of Ce-based `115` compounds \cite{SarraoReview}---it is done by replacing a Ce ion with a non-magnetic one (La). \cite{twofluid1,twofluid2,JPNature} As another possibility, the substitution of magnetic (Yb) ions on the Ce sites provides an extra dimension to the problem:\cite{WilsonRatio,PNAS} since Yb ions are in the mixed valence state of Yb$^{2+}$ (non-magnetic valence configuration) and Yb$^{3+}$ (magnetic valence configuration),\cite{WilsonRatio,Yb-valence1,Yb-valence2} one can study not only how normal and superconducting states evolve as the alloy crosses over from the predominantly local-moment ($x<0.6$) to the mixed-valence ($x>0.6$) regime, but also the problem of the Kondo impurity in the heavy-fermion metal when $x\sim1$. 

Strong hybridization between the conduction and $f$-orbitals  leads to the significant enhancement of the effective mass of the charge carriers at low-temperatures as well as emergence of the effective energy scale $T_F^*$, which is much smaller then the conduction electron Fermi energy. Thus, the onset of unconventional superconductivity at a superconducting transition temperature ($T_c$) around 2.3 K in CeCoIn$_5$ relegates this material to the class of high-temperature superconductors since $T_F^*/T_c\sim20$ only. The origin of such a high (compared to $T_F^*$) transition temperature is thought to be due to system's proximity to a magnetic field induced quantum phase transition, so that strong antiferromagnetic fluctuations provide the source of strong superconducting coupling for the heavy electrons and consequently yield the relatively high value of the superconducting transition temperature.\cite{Gill1,Gill2} 

Recently, the idea of unconventional superconductivity induced by the proximity to a magnetic quantum critical point in CeCoIn$_5$ has been challenged by the systematic study of the superconducting alloys Ce$_{1-x}$Yb$_x$CoIn$_5$.\cite{PNAS} In particular, the evolution of the critical field ($H_{QCP}$) along the $c$-axis at which the quantum phase transition between the antiferromagnetic and paramagnetic phases takes place \cite{expHQCP1,expHQCP2,expHQCP3,expHQCP4,expHQCP5,expHQCP6,expHQCP7} has been studied as a function of Yb concentration. It is interesting to note that the value of $H_{QCP}$ can be extracted from both normal and superconducting states. The authors \cite{PNAS} found that the value of $H_{QCP}$ is significantly reduced for $x>0.10$ and vanishes as $x\to0.20$. Yet, at the same time, the value of the superconducting critical temperature is reduced by only about 15\%  at $x\approx0.20$. Furthermore, the critical temperature is reduced only by half at $x\approx0.5$; this is completely unexpected for an unconventional superconductor given the fact that ytterbium ions are in a mixed valence state with the average valence of $v_{Yb}\approx2.3$,\cite{Yb-valence1,Yb-valence2} so that the magnetic valence state of Yb$^{3+}$ together with essentially unscreened cerium moments should provide enough scattering to completely suppress superconductivity. Thus,  previously reported robustness of the unconventional superconductivity in Ce$_{1-x}$Yb$_x$CoIn$_5$ for intermediate values of Yb concentrations \cite{LiShu} represents a major challenge to our understanding of the physics of this $f$-electron alloy.

In this paper, we report the results of a systematic transport and thermodynamic study of the Ce$_{1-x}$Yb$_x$CoIn$_5$ alloy, mostly focusing on the intermediate range of concentrations, $0.40\leq x\leq0.775$. As we have already mentioned above, the effect of relatively small Yb substitution $(x\leq0.20$) is very dramatic on the physics related to the presence of the magnetic quantum critical point.\cite{PNAS} In the intermediate doping range, however, we observe a crossover from  coherent cerium Kondo lattice behavior to a coherent many-body state formed by the strong hybridization between conduction electrons and ytterbium $f$-electrons.  We note that
the temperature ($T_{coh}$) at which resistivity reaches its peak, referred to as coherence temperature in the text, has a minimum as a function of Yb concentration at $x\sim 0.6$, signaling the reconstruction of the Fermi surface, in agreement with the recent quantum oscillations experiments.\cite{FermiSurface} On the other hand, the monotonic dependence\cite{LiShu} of $T_c$ on $x$ suggests that the onset of many-body coherence in the lattice of Yb ions diluted with Ce $f$-moments is decoupled from unconventional superconductivity. Moreover, our observation of the single impurity Kondo behavior of Ce ions in magneto-transport for $0.40 \leq x \leq 0.70$ points toward an intriguing possibility that unconventional superconductivity has purely local origin and is driven by the presence of Ce $f$-moments. 

\section{Experimental details}

Single crystals of Ce$_{1-x}$Yb$_x$CoIn$_5$ ($0\leq x \leq 0.775$) were grown using an indium self-flux method. The crystal structure and composition were determined from X-ray powder diffraction and energy dispersive X-ray techniques. The single crystals studied have a typical size of $2.1 \times 1.0 \times 0.16$  mm$^3$, with the $c$-axis along the shortest dimension of the crystals. They were etched in concentrated HCl for several hours to remove the indium left on the surface during the growth process and were then rinsed thoroughly in ethanol. 

The electrical transport measurements were carried out on a Quantum Design Physical Properties Measurement System (PPMS) platform. The sample is cooled by flowing cold helium vapor across the sample space.  A heater located near the sample regulates the amount of cooling that is received by the sample, while a thermometer located near the sample allows continuous and automatic control of the sample temperature.  Resistivity measurements were performed using a standard four probe technique. Four electrodes were
attached to the $ab$ surface of the single crystals by bonding Au wires to the crystals with H20E epoxy paste and curing it at $200 ^0$C for 5 minutes. The electrical current $I$ was applied parallel to the $a$-axis of the single crystals and the magnetic field $H$ (up to 14 T) was applied parallel to the $c$-axis of the crystals.

We performed magnetoresistivity measurements under pressure on the $x=0.75$ sample. The single crystal was mounted in a clamped piston-cylinder pressure cell along with a piece of tin wire. A 50:50 mixture of n-pentane and iso-pentane was used as hydrostatic pressure transmitting medium. The superconducting transition temperature of tin was measured in order to determine the actual pressure inside the pressure cell at low temperatures, when the pressure transmiting fluid freezes.

We performed heat capacity measurements on all the single crystals studied for temperatures ($T$) between 1.9 K and 5 K in a Quantum Design PPMS semi-adiabatic calorimeter using a heat-pulse technique. The masses of the single crystals used in these measurements were in the range 1.2 mg to 2.7 mg.

\section{Experimental results}
\subsection{Resistivity}

\begin{figure}
\centering
\includegraphics[width=1.0\linewidth]{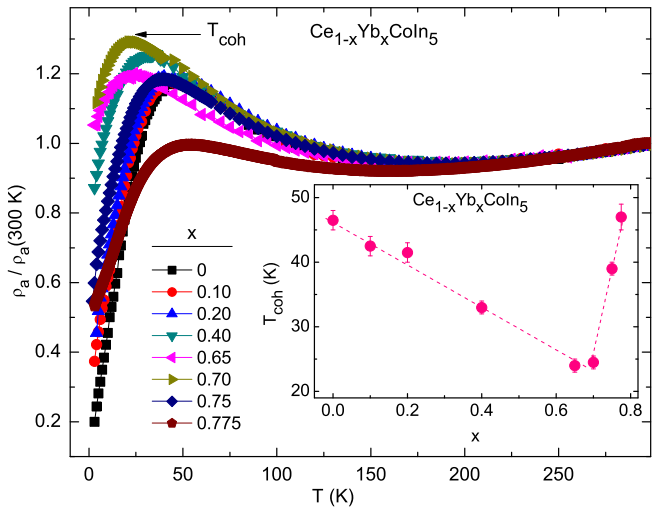}
\caption{(Color online) (a) Temperature $T$ dependence of electrical resistivity $\rho_{a}$ normalized to its value at 300 K for single crystals with different Yb doping ($0 \leq x \leq 0.775$). The arrow on the top marks the coherence temperatures $T_{coh}$, corresponding to the maximum of resistivity. Inset: Coherence temperature $T_{coh}$ vs Yb doping $x$.} 
\label{Fig1}
\end{figure}

Figure \ref{Fig1}(a) shows the $T$ dependence of the electrical resistivity ($\rho_a$) normalized to its value at 300 K of the single 
crystals used in the present study ($0 \leq x \leq 0.775$). In the inset to Fig. \ref{Fig1} we plot the dependence of $T_{coh}$ on Yb doping. We notice that $T_{coh}$ decreases linearly with increasing $x$ up to $x \approx 0.65$ and then it increases with further increasing $x$. This indicates that the cerium Kondo lattice coherence  initially becomes weaker, due to the  dilution of the Ce Kondo lattice with increasing Yb doping, and then the coherence in the system is re-established for Yb doping higher than 65\%. Therefore, this suggests that the coherent many-body state formed by the conduction electrons and cerium $f$-moments is fully destroyed at $x \approx 0.65$ and a new coherent state emerges due to the interaction between the conduction electrons and ytterbium $f$-electrons as if it becomes energetically favorable for the conduction electrons to start hybridizing strongly with ytterbium ions in the lattice. It is also important to note, that our subsequent analysis in section III.C. reveals that cerium ions for the $x = 0.75$ sample are not fully in a single-impurity regime either, while the resistivity decreases with temperature---a clear sign of a coherent metallic state. 
Keeping in mind the results of the recent cyclotron resonance data,\cite{FermiSurface} which show substantial reconstruction of the Fermi surfaces for $x>0.5$, we conclude that the coherence is re-established for $x>0.65$ due to the onset of strong hybridization between the conduction electrons and Yb $f$-electrons, while the Ce moments become decoupled from the conduction band. 

\subsection{Residual resistivity}
At small concentrations $x$ of ytterbium ions, the Kondo lattice coherence temperature appears to be only slowly suppressed with increasing $x$.  This implies that the correlations between Yb impurities play an important role. It has been shown through a combination of x-ray absorption and photoemission spectroscopy techniques\cite{Yb-valence1,Yb-valence2} that the average Yb valence in Ce$_{1-x}$Yb$_x$CoIn$_5$ decreases monotonically from $v_{Yb}=3$ to $v_{Yb}=2.3$ with increasing Yb from $x=0$ to $x = 0.20$ and then it remains constant for $x>0.20$. Therefore, upon the substitution of ytterbium on cerium sites, an ytterbium ion finds itself in a mixed valence state. It is important to keep in mind, however, that the process during which an ytterbium ion changes its valence is a dynamical one. Yb has thirteen $f$-electrons in its magnetic state, while it has fourteen $f$-electrons in its non-magnetic state. Thus, a transition of Yb from its magnetic to non-magnetic state results in an extra $f$-electron and hence an increase in the local volume around Yb. Theoretically, this increase in the local volume is naturally described by introducing the strain dependent hybridization between the conduction and $f$-states of ytterbium.\cite{Maxim} Of course, the conduction electrons may induce Yb-Yb correlations via an analog of the famous Ruderman-Kittel-Kasuya-Yosida (RKKY) mechanism. However, theoretical estimates by Dzero and Huang show that the lattice strain serves as a mediator between two ytterbium ions and leads to an increase in the pair correlation radius between Yb impurities.\cite{Maxim} In addition, they have shown that the presence of Yb-Yb  correlations in a mixed valence alloy must result in a non-linear dependence of the residual resistivity ($\rho_{a0}$) on $x$, with a \emph{ positive quadratic} increase of $\rho_{a0}(x)$,  indicating that the effective interactions between ytterbium ions are predominantly attractive. Also, their theory predicts that the impurity correlations have a healing effect on the formation of the many-body states; i.e., both the Kondo lattice coherence temperature and superconducting critical temperature are expected to decrease at a much lower rate compared to the case when Yb ions are uncorrelated or repel each other. \cite{Maxim} This conclusion is consistent with the slow suppression of coherence and superconductivity\cite{LiShu} in Ce$_{1-x}$Yb$_x$CoIn$_5$ compared with other rare-earth substitutions on the Ce-site.\cite{twofluid2,JPNature} 

\begin{figure}
\centering
\includegraphics[width=1.0\linewidth]{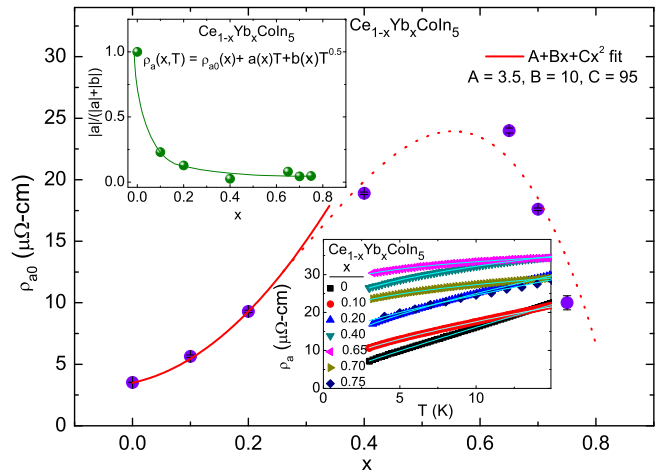}
\caption{(Color online) Plot of residual resistivity $\rho_{a0}(x)$ as a function of Yb doping, for single crystals with $0 \leq x \leq 0.75$. The error bars of $\rho_{a0}(x)$ are the standard errors given by the fitting routine and they are within the symbols size. The solid line is a quadratic polynomial fit of the residual resistivity data at low-$x$ values, while the dotted curve is a guide to the eye for the higher $x$ values. Right Inset:  Fit of temperature $T$ dependent resistivity data (as discussed in the text). Left inset: Doping dependence of the linear-in-temperature contribution divided by the total contribution to resistivity, obtained from fits of the resistivity data (right inset) for $0 \leq x \leq 0.75$ (the fit of resistivity data is in accordance with Eq. (1) discussed in the text). }
\label{Fig2}
\end{figure} 

To further test these ideas, we studied the dependence of  $\rho_{a0} \equiv \rho_a(0$ K) on $x$. We extracted $\rho_{a0}$ of all the single crystals studied by fitting $\rho_{a}(T)$ in the temperature range 3 K $\leq T \leq 15$ K with
\begin{equation}\label{ResistivityCe}
\rho_a(x,T) = \rho_{a0}(x) + a(x)T + b(x) \sqrt T.
\end{equation}  
The right inset to Fig. \ref{Fig2} shows the excellent agreement between the $\rho_{a}(T)$ data and the fitting curves. 

\begin{figure}
\centering
\includegraphics[width=1.0\linewidth]{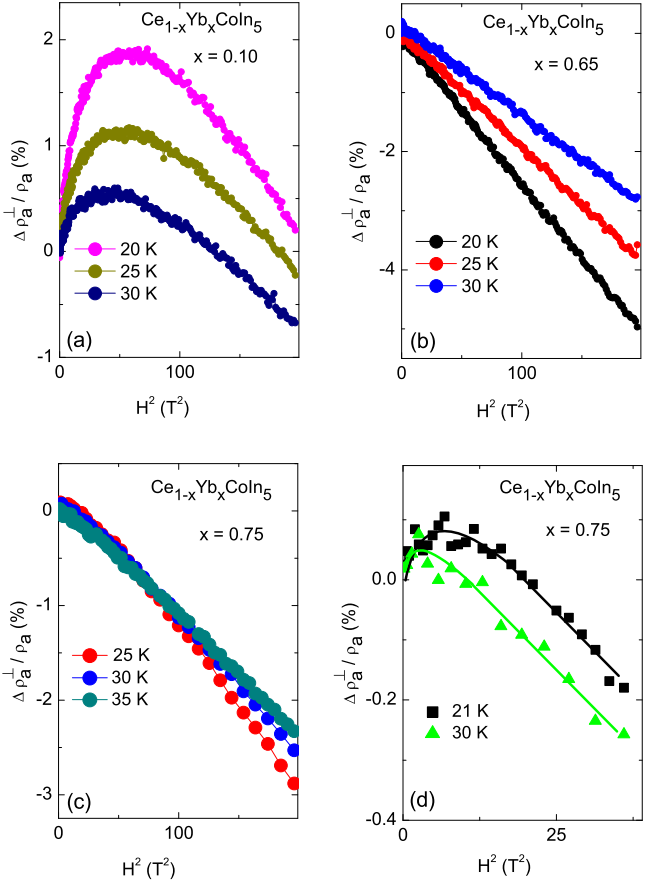}
\caption{(Color online) Magnetoresistivity $\Delta \rho_{a}^{\perp}/\rho_a$ as a function of magnetic field squared $H{^2}$ for the (a) $x = 0.10$, (b) $x = 0.65$, (c) $x = 0.75$ single crystals, and for (d) $x=0.75$  single crystals for two temperature and low field values. }
\label{Fig3}
\end{figure}

The main panel of Fig. \ref{Fig2} shows the results of $\rho_{a0}(x)$. The solid line is a quadratic polynomial fit of the residual resistivity data at low-$x$ values, with a {\it positive} quadratic term, while the dotted curve is a guide to the eye for the higher $x$ values. This result  is in excellent agreement with the theory of Dzero and Huang,\cite{Maxim} indicating that, indeed, the effective interactions between ytterbium ions are predominantly attractive. We note that the non-monotonic $\rho_{a0}(x)$ is expected since the increase in $\rho_{a0}(x)$ and the corresponding maximum for $x\approx 0.55$ appear as a consequence of introducing disorder by the Yb substitution into the ordered host. 

Lastly, we note that the reason for choosing the temperature dependence given by Eq. (1) was discussed in details elsewhere.\cite{PNAS} Here we want to mention that the presence of critical fluctuations at lower Yb doping ($x \leq 0.20$) gives rise to the linear $T$-dependent contribution, while the $\sqrt T$ contribution is attributed to the cooperative mixed valence state of Yb. The left inset to Fig. 2 gives the ratio of the $T$-linear contribution to the total $T$-dependent contribution of the resistivity, i.e.,  $|a|/(|a| + |b|)$, which reflects the percentage of the linear in $T$ term. This inset shows that the linear-in-$T$ contribution decreases with increasing $x$ and become almost zero for $x \geq 0.20$. This indicates that, indeed, the system is predominately in the quantum-critical regime only up to $x \approx 0.20$. 

\begin{figure}
\centering
\includegraphics[width=1.0\linewidth]{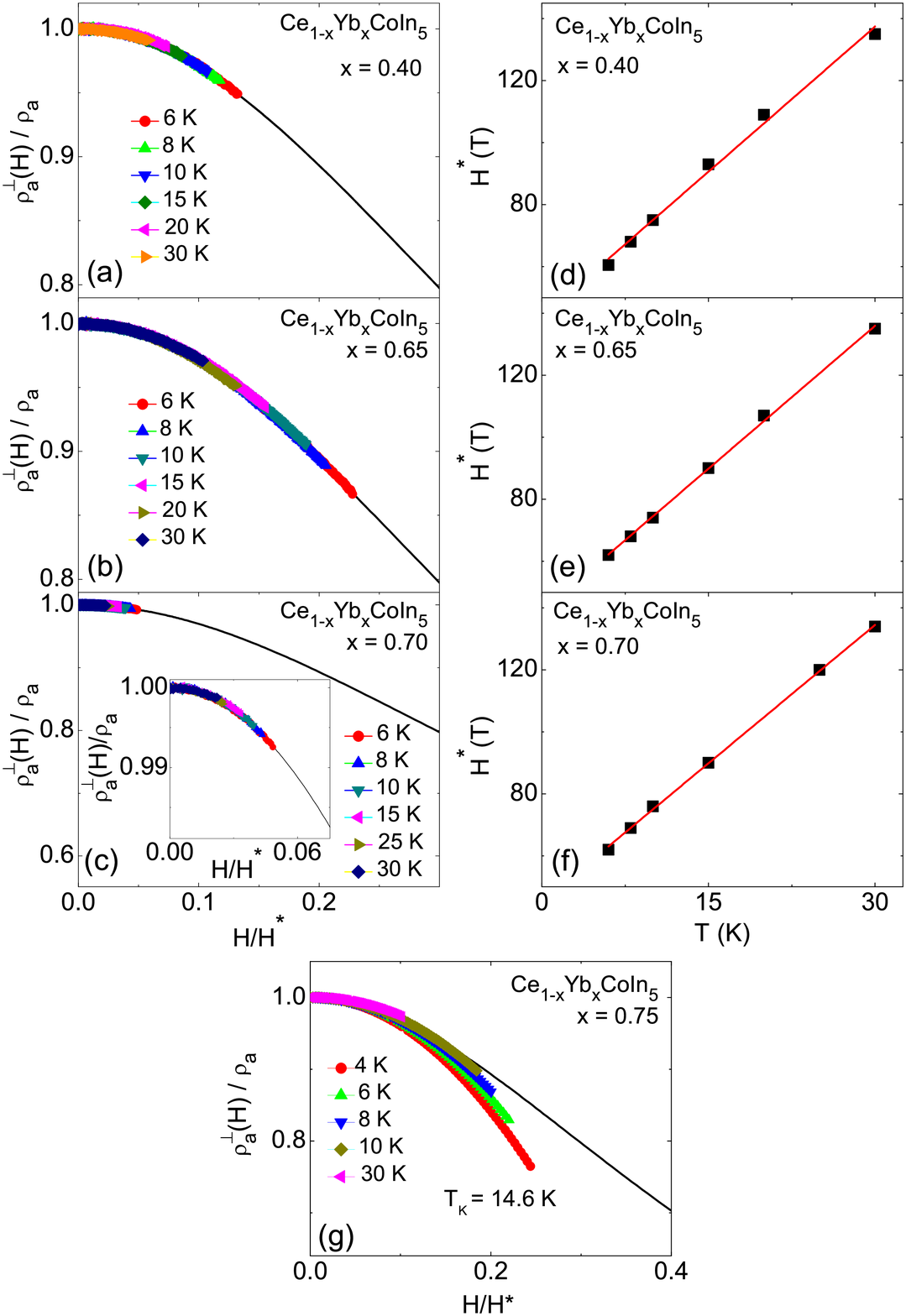}
\caption{(Color online) (a), (b), (c), and (g): Plot of normalized magnetoresistance $\rho_{a}^{\perp}(H)/\rho_a$ data vs $H/H^*$ for the single crystals in the doping range $0.40 \leq x \leq 0.75$, measured at different temperatures $T$. Inset to (c): Same data as in the main panel zoomed at lower $H$ values. The solid curves are the Schlottmann's theoretical magnetoresistance curve. The scaling field $H^*$ has been adjusted at each temperature to optimize the scaling of the experimental data and Schlottmann's theoretical magnetoresistance curve. Panel (g) shows that Schlottmann scaling does not work for $x \geq 0.75$. The panels (d), (e) and (f) show scaling parameter $H^*$ vs $T$ for all three doping for which Schlottmann's scaling works. }
\label{Fig4} 
\end{figure}

\subsection{Magnetoresistivity}

Magneto-transport measurements are an excellent tool to probe the physics of heavy-fermion materials; they can provide insight about many-body coherence, single-ion Kondo behavior, and quantum criticality.\cite{PNAS}  In order to further verify the ideas developed while performing the in-plane resistivity measurements, we performed transverse ($H \perp ab$) in-plane magnetoresistivity (MR), $\Delta \rho_{a}^{\perp}/\rho_a \equiv [\rho_a^{\perp}(H)- \rho_a(0)]/\rho_a(0)$, measurements as a function of $T$ and applied magnetic field ($H$). The data for all the doping studied fall under three groups: (i) a positive contribution to MR for low $H$ values  followed by a negative $H^2$ dependence at higher $H$ values for $x \leq 0.20$ [Fig. \ref{Fig3}(a)], (ii) a negative and quadratic in $H$ contribution to MR, typical of a dilute Kondo system, for the intermediate doping range, $0.20 < x \leq 0.70$, [Fig. \ref{Fig3}(b)], and (iii) the re-appearance of the positive contribution to MR  for $x \geq 0.75$ [Figs. \ref{Fig3}(c) and \ref{Fig3}(d)]. 

The positive  in-plane magnetoresistivity (MR) is a result of antiferromagnetic critical spin fluctuations\cite{PNAS} or, according to the standard theory of Kondo lattice systems,\cite{PiersReview} a result of the formation of the coherent Kondo lattice state,\cite{Positive1,Positive2,Positive3} in heavy fermion materials measured at $T<T_{coh}$ and at low $H$ values. Therefore, the results of Figs. \ref{Fig3} are consistent with the non-monotonic $T_{coh}(x)$ dependence (inset to Fig. 1) and confirm that the coherence effects are stronger in low- and high-Yb-doped samples, while incoherent scattering dominates the contribution to MR for the intermediate Yb doping range. 

A further analysis of the magnetoresistivity data based on Schlottmann scaling \cite{SingleImpurity} confirms the above findings. Specifically, with increasing Yb concentration, one expects to observe the  crossover from the cerium Kondo lattice to cerium Kondo impurity behavior. 
In the single-impurity Kondo limit, a plot of the $\rho_a^{\perp}(H)/\rho_a$ data vs $H/H^*$, where  
\begin{equation}
H^*=k_B(T+T_K)/g\mu_B
\end{equation}
and $T_K$ is the single impurity Kondo temperature, must display Schlottmann scaling.\cite{SingleImpurity}  This universal scaling shows that the underlying physics of the Kondo impurity in the single-ion regime is dominated by a single energy scale related to $T_{K}$. Figures \ref{Fig4}(a), \ref{Fig4}(b), and \ref{Fig4}(c) show the results of this universal scaling for the $x = 0.40$, 0.65 and 0.70 samples, respectively, for 4 K $\leq T \leq 30$ K. The Schlottmann curve (solid lines in Figs. \ref{Fig4}) on which the experimental data scale is for $S=1/2$ case. Choosing $S=1/2$ is appropriate in the present case as Ce$^{3+}$ follows the Kondo impurity model for $J=5/2$ ion under the tetragonal crystalline field.\cite{twofluid2} Notice that the scaling works very well for all these Yb concentrations, with a linear dependence of $H^*$ on $T$ [Figures \ref{Fig4}(d), \ref{Fig4}(e), and \ref{Fig4}(f)], as expected based on Eq. (2), confirming that, indeed, the system is in the single Ce Kondo impurity limit for this intermediate Yb doping range. 
The Kondo temperature, obtained from this linear relationship [see Eq. (2)], has the {\it same} value $T{_K}= 14.6\pm 0.4$ K for these three Yb doping. The temperature scale of 14  K is most likely an intermediate energy scale separating the purely single ion behavior of cerium with $T_K \approx 2$ K and purely coherent behavior of cerium Kondo lattice with $T_{coh} \approx 45$ K. In other words, it reflects some degree of hybridization with conduction electrons as well as the antiferromagnetic exchange interaction between the Ce ions. 

Schlottmann's scaling does not work for any $H^*$ value for $x \geq 0.75$. Figure \ref{Fig4}(g) shows an example with $H^*$ corresponding to $T_K=14.6$ K. This lack of Schlottmann's scaling indicates that the Ce ions are not fully in a single impurity regime and that the dominant contribution to scattering for these higher doping is coming from the itinerant Yb $f$-electronic states strongly hybridized with the conduction electrons.

\begin{figure}
\centering
\includegraphics[width=1.0\linewidth]{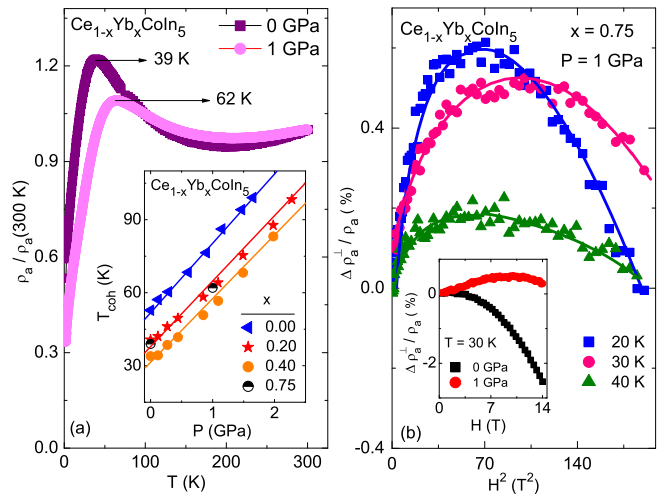}
\caption{\label{Fig 5}(Color online)(a) Temperature $T$ dependence of normalized electrical resistivity $\rho{_a}/\rho{_a} (300$ K) for the $x = 0.75$ single crystal measured at ambient pressure and at 1 GPa. Inset: Pressure $P$ dependence of coherence temperature $T_{coh}$. The data for $x=$ 0.00, 0.20, and 0.40 samples are taken from Ref. \onlinecite{PressureYb}. The data for the $x = 0.75$ sample are from the present work. (b) Magnetoresistivity $\Delta \rho_{a}^{\perp}/\rho_a$ of the $x = 0.75$ single crystal measured at different temperatures and under 1 GPa. Inset: Comparison of magnetoresistivity $\Delta \rho_{a}^{\perp}/\rho_a$ of the $x = 0.75$ single crystal measured at 30 K in ambient pressure and at 1 GPa.}
\label{Fig5}
\end{figure}

\begin{figure}
\centering
\includegraphics[width=1.0\linewidth]{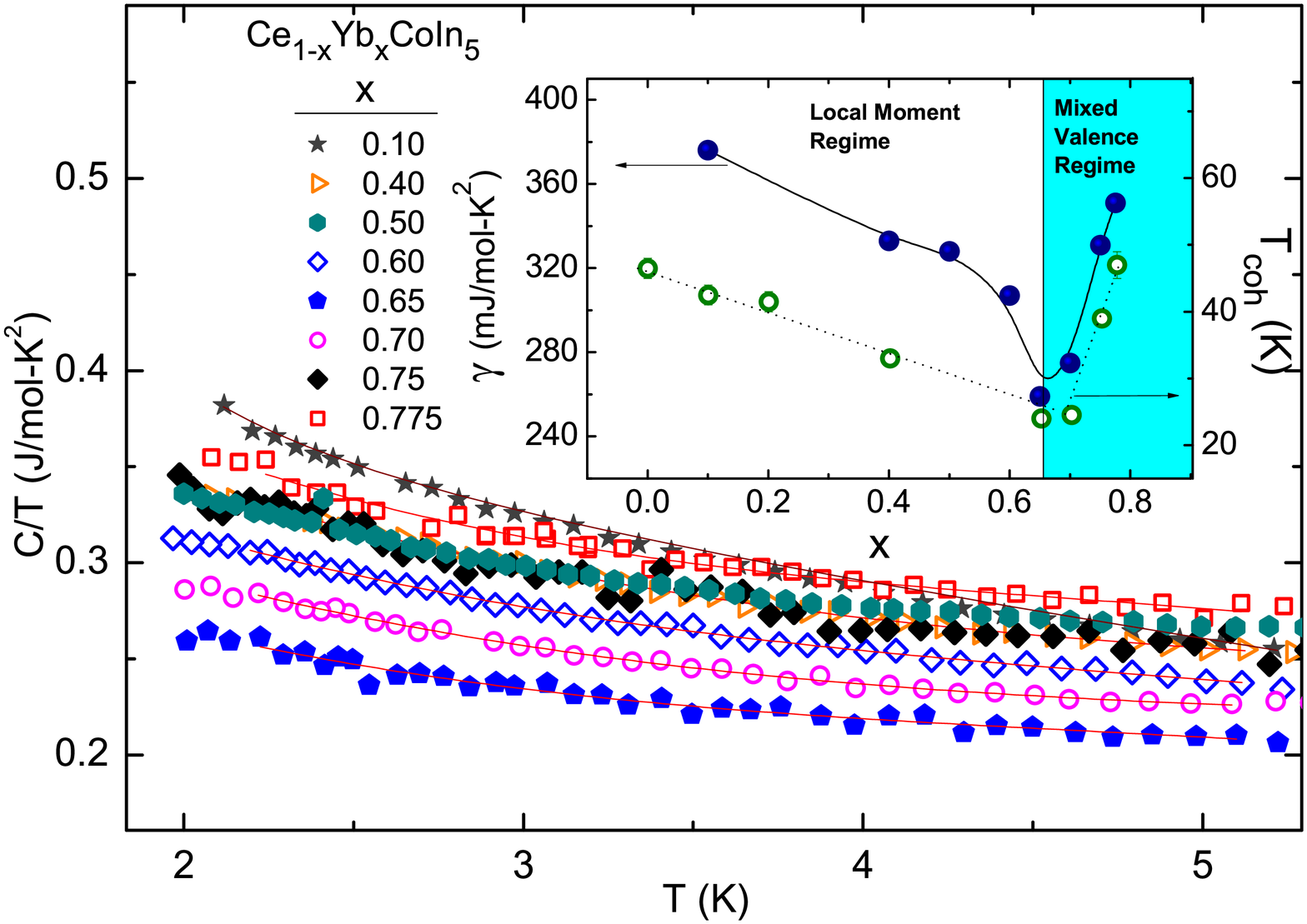}
\caption{(Color online) Specific heat $C$ divided by temperature $T$ for single crystals with different Yb concentrations  ($0.10 \leq x \leq 0.775$), measured over the temperature range 1.9 to 5 K. Inset: Value of Sommerfeld coefficient $\gamma$ vs $x$ for $T = 2.15$ K and $T_{coh}$ vs $x$. The Ce and Yb sides are delineated  on the figure as the local moment regime and the mixed valence regime, respectively.}
\label{Fig6}
\end{figure}

\subsection{Magnetoresistivity under pressure}

As discussed above, for $x \geq 0.75$, the scaling for the single-ion Kondo behavior of Ce moments does not work, $T_{coh}(x)$ increases with increasing $x$, and the positive contribution to MR reappears in  this higher doping range. All these results reflect the strengthening of coherence in the high Yb doping range. Recent results on Ce$_{1-x}$Yb$_x$CoIn$_5$ [solid symbols in the inset to Fig. \ref{Fig5}(a)] have shown that $T_{coh}$ increases linearly with pressure ($P$) with the same slope for different Yb doping. \cite{PressureYb} 

To study the effect of pressure on the coherence developing at high Yb doping, we performed MR measurements under pressure on the $x=0.75$ sample. Figure \ref{Fig5}(a) shows that an applied pressure of 1 GPa increases the coherence temperature from 39 K to 62 K [half filled symbols in the inset to Fig. \ref{Fig5}(a)]. This increase is {\it quantitatively} in agreement with the recently published results.\cite{PressureYb} Figure \ref{Fig5}(b) shows the MR data measured under 1 GPa  and at different temperatures. The positive contribution to MR clearly becomes stronger under pressure [inset to Fig. \ref{Fig5}(b)] for all the temperatures, which indicates that pressure increases coherence.

\subsection{Heat capacity}

The previously discussed transport measurements show that there is a crossover between Kondo lattice behavior driven by the hybridization between the conduction band and cerium $f-$moments to coherent state governed by the hybridization between the conduction states and ytterbium $f-$orbitals for Yb concentrations larger than $65 \%$. As a final test of this finding, we performed heat capacity measurements on all the single crystals studied. Figure \ref{Fig6} shows specific heat ($C$) data divided by temperature measured for 1.9 K $\leq T \leq$ 5 K. The value of the Sommerfeld coefficient $\gamma \equiv C/T $ obtained at $2.15$ K is shown in the inset. All Yb doping levels give large values of $\gamma$ (260 mJ/mol$\cdot$K$^2 \leq \gamma \leq$ 380 mJ/mol$\cdot$K$^2$), which confirms the heavy fermion nature and strong electronic correlations for all Yb concentrations. In addition, $\gamma(x)$ and $T_{coh}(x)$ show the same behavior [inset to Fig. \ref{Fig6}], i.e., they decrease monotonically with increasing Yb concentration up to $x = 0.65$ and then they increase with further increasing $x$. This behavior of $\gamma(x)$ reflects the fact that the quasiparticles are stabilized easily with strengthening coherence in the system at higher Yb doping, which in turn yields a larger effective mass. This shows that the coherence effects, indeed, are strengthened for $x > 0.65$.

\section{Conclusions}

In this paper we have reported transport and thermodynamic studies 
of the heavy-fermion superconducting alloys Ce$_{1-x}$Yb$_x$CoIn$_5$ focusing on the intermediate range of concentrations, i.e., $0.40\leq x\leq0.775$. We observed a  crossover from the local moment regime, driven by the hybridization between the conduction electrons and cerium $f$-moments, to a coherent many-body state formed by the strong hybridization between conduction electrons and ytterbium $f$-electrons. The temperature $T_{coh}$ at which resistivity reaches its peak has a minimum as a function of Yb concentration at $x\approx 0.65$, while $T_c$ decreases linearly with $x$ over the whole range.\cite{LiShu}
This implies that the onset of the many-body coherence in the lattice of Yb ions diluted with Ce $f$-moments and unconventional superconductivity are decoupled from each other. The signature of the Ce single-impurity Kondo behavior is revealed by the magneto-transport data for $0.4 \leq x \leq 0.70$. This suggests the likelihood that unconventional superconductivity has purely local origin and is driven by the presence of cerium $f$-moments.
Our present results as well as previous transport, thermodynamic, and spectroscopic results show a systematic change in the properties
of Ce$_{1-x}$Yb$_x$CoIn$_5$ for $0\leq x<0.8$. The fact that superconductivity exists in alloys with $x\sim 0.8$ clearly implies that superconductivity is driven by the presence of Ce ions as though they effectively act as negative-$U$ centers. 
Given the complete suppression of the antiferromagnetic fluctuations for $x>0.20$,\cite{PNAS} superconductivity may either be mediated by the 
electron-phonon interaction \cite{BCS} or may be due to a more unconventional mechanism involving virtual fluctuations into higher lying Ce crystalline field multiplets.\cite{Flint}  

This work was supported by the National Science Foundation (grant NSF DMR-1006606) and Ohio Board of Regents (grant OBR-RIP-220573) at KSU, and by the U.S. Department of Energy (grant DE-FG02-04ER46105) at UCSD. 
\\

\end{document}